# A Machine Learning Based Algorithm for Joint Improvement of Power Control, link adaptation, and Capacity in Beyond 5G Communication systems

Jafar Norolahi, Paeiz Azmi
Department of Electrical and Computer Engineering
Tarbiat Modares University
Tehran, Iran
PAZMI@modares.ac.ir

*Abstract*— In this study, we propose a novel machine learning based algorithm to improve the performance of beyond 5 generation (B5G) wireless communication system that is assisted by Orthogonal Frequency Division Multiplexing (OFDM) and Non-Orthogonal Multiple Access (NOMA) techniques. The non-linear soft margin support vector machine (SVM) problem is used to provide an automatic modulation classifier (AMC) and a signal power to noise and interference ratio (SINR) estimator. The estimation results of AMC and SINR are used to reassign the modulation type, codding rate, and transmit power through frames of eNode B connections. The AMC success rate versus SINR, total power consuming, and sum capacity are evaluated for OFDM-NOMA assisted 5G system. Results show improvement of success rate compared of some published method. Furthermore, the algorithm directly computes SINR after signal is detected by successive interference cancellation (SIC) and before any signal decoding. Moreover, because of the direct sense of physical channel, the presented algorithm can discount occupied symbols (overhead signaling) for channel quality information (CQI) in network communication signaling. The results also prove that the proposed algorithm reduces the total power consumption and increases the sum capacity through the eNode B connections. Simulation results in compare to other algorithms show more successful AMC, efficient SINR estimator, easier practical implantation, less overhead signaling, less power consumption, and more capacity achievement.

*Keywords*- Beyond 5G, OFDM, NOMA, Machine Learning, Support Vector Machine, Automatic Modulation Classification, SINR Estimation, Power Control, Link Adaptation, Network Capacity Improvement.

## I. INTRODUCTION

Today, societies are extremely related to communication networks for their demands. Because of increasing daily demands, the communication networks are daily becoming enormous and already are upgrading their generations with new technologies. For instance, OFDM and NOMA are approaches to resolve the growing enormous density of communication networks.

The communication networks, being responsible and willing to huge user equipment (UE) in automotive and intelligent communication, must be optimally characterized by their dynamic, heterogeneous, complex, and intensive nature. Recently, machine learning (ML) is attended to improve and optimize the communication networks [1]. Since the ML are more performing than other approaches, they are highly appropriate to characterize the mobile communication networking architecture and dynamic environment. Now, ML is one of the significant issues in beyond $5^{th}$ generation (B5G) wireless mobile networks to improve power control, link adaptation, capacity, scheduling and routing of traffic, and coverage. It means that the stringent requirements and highly dynamic conditions of B5G should be considered [2]. The 5G, supporting the high-density network users, has at least three serious challenges in power consumption, link adaptation, and capacity. Firstly, without appropriate power control for power consumption, low quality of service (QoS) and saturation of receivers in eNode B can happen. In addition, power control can be useful to suppress the uplink interferences because increasing number of UEs can lead to intensify the congested signals in eNode B. So, a proper power control strategy is needed to manage the uplink interferences and influence the QoS among the UEs who select the same sub channel in the OFDM-NOMA system.

On the other hand, in OFDM-NOMA system, a very heavy signaling overhead is imposed by hierarchical cancellation structure. For example, the overhead increases with a factor ($M + 1$)/2 in NOMA which $M$ *is* the number of users superposed onto the same resources. Since the overhead signaling consists of information about the modulation and channel coding (link adaptation) of different users and their allocated powers, suppression of the overhead signaling can improve spectral efficiency [4]. For more illustration, in the case of overhead signaling, the useful symbol parts for Long-Term Evolution (LTE) are changed from $14 - 3 = 11$ symbols to $14 - 4 = 10$ symbols forcing a decrease of 10% in the spectral efficiency [4]. It means that near the 28% of symbols are occupied by the information about the 3 parameters, modulation type, channel coding of different users, and their allocated power to overcome the channel conditions. Therefore, when the number of users grows in 5G, the occupied capacity by the overhead signaling is extremely noticed by researchers. In addition, in [5]-[9] some algorithms of real time automatic modulation classification (AMC) are reported for B5G. The receivers of B5G system and unmanned aerial vehicle (UAV), receiving signal from multiple sources and directions, experience deep multipath fading and difficult identifying of signals. Thus, for having a successful link adaptation, the receivers should know physical channel conditions and the types of modulation-coding couple of users in each frame.

Additionally, literatures categorize the AMCs into two general methods: likelihood-based (LB) methods and feature-based (FB) methods [10]-[13]. The LB methods compare the likelihood ratio of each possible hypothesis against a threshold which is derived from the probability density function of the achieved signal. Multiple likelihood ratio test (LRT) algorithms have been proposed as Average LRT [14], Generalized LRT [15], and quasi-hybrid LRT [16].

The ability of modulation detection in low SNR conditions using short observation intervals is one of most important criteria of good classifiers [17]- [20]. These classifiers must also be robust against processing errors and be able to detect large number of modulations in different propagation conditions. Besides, real-time operation and low computational complexity have been considered in [21]-[25].

In some researches, AMCs are offered for discounting the overhead signalling which are suitable for capacity improvement. For example, what are reported in [26]- [38] are novel methods that identify the modulation constellation of different users via feature-based AMC algorithms before symbol detection. This means that the network does not need to occupy some symbols to notify each user about modulation and channel codding type. As a result, proposing a novel method to decrease the number of occupied symbols is necessary.

Another parameter that is widely utilized for power control and capacity management is SINR estimation [39]- [47]. SINR estimation is conveyed by many researches in different methods. For instance, in [48], a method based on Gray model is proposed. It is more appropriate for macro-pico network with strong correlated channels. Also, SINR estimation is investigated in LTE and OFDM based network in [49] and [50] by the aid of ratio of the transmitted signal power to the error signal power that it is measured after decoding and error measurement.

As an another result, using machine learning approach in our study makes available a novel idea to SINR estimation before decoding.

Therefore, we also propose machine learning based algorithm for joint AMC and SINR estimation to decrease power wasting and suppress overhead signaling. The joint AMC and SINR estimation are performed by the aid of nonlinear soft margin support vector machine (SVM) problem. The SVM results provide a capability to reassigning modulation type, coding scheme, and transmit power without occupied symbols to prior channel quality information (CQI). SINR computation is founded to ratio of the transmitted signal power to the error signal power where symbols of error are estimated by support vectors in SVM margin. In this situation, communication network can implement power control, link adaptation, and capacity enhancement with less overhead signaling.

This paper is organized as following. After introducing the suggested algorithm and its motivation in part I, the related works that they studied the ML base issues in AMC, power control, link adaptation, and capacity improvement are noticed in part II. The mathematic basis and data modelling of the proposed method are informed in part III. The part III includes five subsections. In first subsection the data modelling for a OFDM-NOMA base system is illustrated. ThenML assisted link adaptation and power control model are presented in second subsection. Next, AMC mathematical model based on related mathematic of nonlinear soft margin SVM problem is described in third subsection. Later, the SINR using estimated support vectors in the SVM margin is calculated in fourth subsection. In last subsection of part III, codding rate of each signal is estimated by the aid of estimated SINR and given parameters such as spectral cfieciency and bitper symbol in a looked-up table. Moreover, in part IV, we present how proposed algorithm work to achieve power control, link adaptation, and capacity enhancement. Furthermore, the simulation results and their analyses are described in part V that it involves AMC success rate in triple regular modulation types in 5G networks, total power consumption of proposed algorithm in compare of current algorithm, and capacity calculation and its mathematic calculate description. Finally, in last part, the conclusion of this study is explained.

## II. Related Works

Machine learning base study in power control and link adaptation is always interesting to researchers. Studies covered wide issues in machine learning. For example, AMC based on deep convolutional neural networks was proposed in [51]. It can improve AMC performance at low SINRs. Furthermore, a deep learning-based algorithm to classify and mitigate jam signals through power control was studied in [52].

Also, in [53], a real time learning algorithm for link adaptation in multiple-input multiple-output (MIMO) systems is implemented. It has an acceptable performance and more quick adaptation in compare to other methods. However, it minimized spatial overhead.

In addition, a Q-learning framework for power control and interference avoidance was proposed in [54]. The interferences of femtocells in eNode B were mitigated while their decentralized self-organization were founded. The Q-learning was also used in [55] that it focused on power control in a multi-cell indoors environment. It uses a robust scheduling in to establish a dedicated channel that prepared the power control of the downlink to voice quality improvement.

Additionally, power control algorithm to 5G was reported in [56]. It focused on moderating of overhead signaling to reduce CQI through the eNode B connections.

For more discussion, deep learning-based algorithm in wireless communications has been studied in [52] and [57]. The deep reinforcement learning to 5G power control was studied in [57]. It improved non-line of sight (NLOS) transmission performance by beamforming and maximized sum-rate of UEs under the constraints of transmission power and quality targets was solved using deep reinforcement learning. An algorithm to maximize the multichannel access via deep Q-learning was done in [58]. Many machine learning assisted classifiers were proposed in other studies. For instance, linear discriminant analysis classifier [59], parallel general regression neural network, and 1-dimentional convolutional neural networks [60] are noticed. There are also Block-Based Neural Networks [61], linear discriminant classifier with feature extracted from discrete wavelet transform [37], and SVM classifiers [62]–[65].

However, most interesting classifier is SVM due to its outstanding advantages [66]. One of SVM advantages is the ability of kernel function utilization because it leads to less cost in practical implementation. For instance, in [66], a practical SVM classifier was proposed. Its results show that the classifier with kernel function of radical basis function (RBF) costs more than 50000 times power consumption per classification compared to a linear kernel function. Likewise, the local and global SVM classification were conveyed by the aid of LIBSVM library [67].

### III. MATHEMATIC BASIS AND DATA MODELLING OF THE PROPOSED METHOD

In this paper, we assume a flat fading multipath channel in urban environment. The network communication uses an OFDM system that is assisted by NOMA scheme in each sub channel [68]. In follow, the required fundamentals of suggested algorithm including four subsections are informed.

*A. Data modeling*

For data modelling, we assume a $N$–user NOMA system in an OFDM system with F sub channels. So, it can be supposed that an eNode B which supports $F \times N$ users. Assume that $[S_1, S_2, \ldots, S_N]$ are the symbols of N different users in each sub channel. Each user can choose its modulation type between different modulation schemes. So, we have $[M_1, M_2, \ldots, M_N]$ where $M_n \in \{1, \ldots, T\}$ represents the label of $T$ different possible modulation schemes for any user. In this study we assumed three modulation schemes, QAM, 16PSK, and 64PSK that have been using in LTE and proposed 5G. It means that $T = 3$. The signal based on OFDM and NOMA can be modeled as follow:

$$S = \sum_{f=1}^{F} \sum_{n=1}^{N} h_{f,n} p_n s_{f,n} + \mathcal{N} \quad (1)$$

where $h_{f,n}$, $p_n$, $s_{f,n}$, and $\mathcal{N}$ are the flat fading Rayleigh channel coefficient of $n^{th}$ user in $f^{th}$ sub channel, the allocated power to $n^{th}$ user in NOMA, modulated symbol of $n^{th}$ user in $f^{th}$ sub channel, and the white Gaussian noise with variance $\sigma^2$, receptively. In Fig.1, proposed method is indicated for $N = 2$. According to Fig.1, $f^{th}$ sub channel of OFDM enters to a Two-User NOMA (TU NOMA). In TU NOMA, first and second signal are separated and are arrived to suggested power control method unit. As mentioned, the suggested method utilizes the ML approach assisted algorithm to provide properly modulation, codding rate, and transmit power to each user. On other words, before achieving power control and link adaptation, some processes must be performed that consist of SVM calculation, AMC, and SINR estimation. The recognized parameters will be used by the suggested algorithm (Algorithm 1 in partIII) to reassigning the modulation and codding rate, and allocating the power transmit in next frame of eNode B connection.

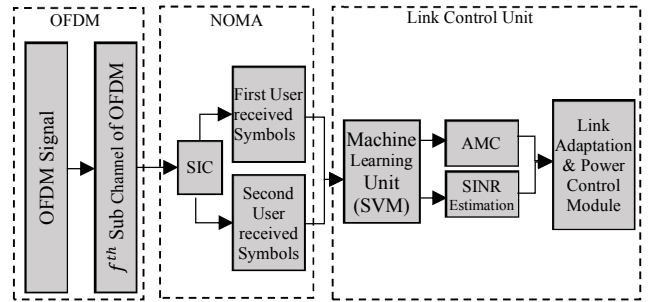

Fig. 1: proposed method for TU-NOMA

In following section, we present that how suggested method estimates jointly modulation and SINR of each user in NOMA.

*B. AMC mathematical model*

Modulation classification problem is confronted with a noisy data that leads to overlapping classes. In this situation, there are nonlinear separable data and nonlinear decision surfaces. In the case of noisy data and overlapping classes of samples, the soft margin SVM, having slack variables in the problem, is more applicable than finding only maximum margin in hard margin SVM. So, we proposed a solution in this study suggests a soft margin SVM to handle classification of nonlinear separable data.

Assume a set of modulated data examples (for example QPSK) along $X \in \mathbb{R}^d$ with their corresponding labels that we saved in $Y = \{y_1, \ldots, y_n\}$. For instance, the QPSK modulated data has 4 classes data where the labels can be chosen from $y_n \in \{c_1, c_2, c_3, c_4\}$. Given a training data set consisting of example–label pairs, $\{(x_1, y_1), \ldots (x_N, y_N)\}$, we would like to estimate parameters of the model that will give the smallest classification error. In this situation, according to [69], optimization problem of soft margin SVM can assumed to:

$$\min_{w, w_0, \{\xi_n\}_{n=1}^{N}} \frac{1}{2} \|w\|^2 + c \sum_{n=1}^{N} \xi_n$$

$$s.t. \quad y^{(n)}(w^T x^{(n)} + w_0) \geq 1 - \xi_n \quad n = 1, \ldots, N \quad (2)$$

$$\xi_n \geq 0$$

where $w$ is a normal vector of the hyperplane, $w_0$ is the intercept, and $\xi_n$ is slack variable corresponding to each example–label pair $(x_n, y_n)$ that allows a particular example to be within the margin or even on the wrong side of the hyperplane. In equation (2), We subtract the value of $c \sum_{n=1}^{N} \xi_n$ from the margin, constraining $\xi_n$ to be non-negative. The parameter of $c > 0$ trades off the size of the margin against the total amount of slack that we have. This parameter is called the "regularization parameter," and the term of $c \sum_{n=1}^{N} \xi_n$ in the objective equation (2) is a regularization term.

Assume a transformation $\phi: \mathbb{R}^d \rightarrow \mathbb{R}^m$ on the feature space. So, we have $x \rightarrow \phi(x)$ and $\phi(x) = [\phi_1(x), \ldots \phi_m(x)]$ where $\{\phi_1(x), \ldots \phi_m(x)\}$ is a set of basic functions or features. By substituting of $\phi(x)$ in (1),

$$\min_{w, w_0, \{\xi_n\}_{n=1}^N} \frac{1}{2} \|w\|^2 + c \sum_{n=1}^N \xi_n$$

$$s.t. \quad y^{(n)}(w^T \phi(x^{(n)}) + w_0) \geq 1 - \xi_n \quad n = 1, \ldots, N \quad (3)$$

$$\xi_n \geq 0$$

where $w \in \mathbb{R}^m$ is the weight set that must be estimated, and in the case of $m \gg d$ (very high dimensional feature space), there are many more parameters to learn. The equation (3) is known as the primal SVM problem. Considering to inputs that $x \in \mathbb{R}^d$ with $d$ features, and $w$ has same dimension as $x$, the number of parameters (the dimension of $w$) in the optimization problem grows linearly with the number of features.

In the following, we consider an equivalent optimization problem (the so-called dual view), which is independent of the number of features. It means that the number of parameters increases with the number of examples in the training set instead of number of features. This is useful for problems where we have more features than the number of examples in the training dataset. The dual SVM also has the additional advantage that it easily allows kernels to be applied [69]. So, applying proper kernel can decrease cost of practical implementation. Corresponding to the primal SVM, We call the variables $w$, $w_0$, and $\xi$ as primal variables.

In addition, we use $\alpha_n \geq 0$ as the Lagrange multiplier corresponding to the constraint (3) that the examples are classified correctly, and $\xi_n \geq 0$ as the Lagrange multiplier corresponding to the non-negativity constraint of the slack variable; see (3). The Lagrangian is then given by

$$L(w, b, \xi, \alpha, \gamma) = \frac{1}{2} \|w\|^2 + c \sum_{n=1}^N \xi_n$$
$$- \sum_{n=1}^N \alpha_n (y_n(\langle w, x_n \rangle + b) - 1 + \xi_n) - \sum_{n=1}^N \gamma_n \xi_n \quad (4)$$

By differentiating the Lagrangian (4) with respect to the three primal variables $w$, $w_0$, and $\xi$ respectively, we obtain

$$\frac{\partial L}{\partial w} = w^T - \sum_{n=1}^N \alpha_n y_n x_n^T$$

$$\frac{\partial L}{\partial b} = \sum_{n=1}^N \alpha_n y_n \quad (5,6,7)$$

$$\frac{\partial L}{\partial \xi_n} = c - \alpha_n - \gamma_n$$

substituting results of (5), (6), and (7) in (4), the optimization problem of Soft-margin SVM in a transformed space (duality problem) is obtained as follow:

$$Max_\alpha \left\{ \sum_{n=1}^N \alpha_n - \frac{1}{2} \sum_{n=1}^N \sum_{m=1}^M \alpha_n \alpha_m y^{(n)} y^{(m)} \phi(x^{(n)})^T \phi(x^{(m)}) \right\} \quad (8)$$

Subject to $\quad \sum_{m=1}^N \alpha_n y^{(n)} = 0$

$\quad 0 \leq \alpha_n \leq c \quad n = 1, \ldots, N$

The (8) equation can be solved by quadratic programming. Additionally, we have an inner products of $\phi(x^{(n)})^T \phi(x^{(m)})$ that it only needs to learn $\alpha = [\alpha_1, \ldots, \alpha_n]$ for achieving the result. In other words, it is not necessary to learn $m$ parameters as opposed to the primal problem. Because of that $\phi(x^{(m)})$ is maybe non-linear function, it is possible to provide a non-linear classifier in examples of $x_n$.

Since $\phi(x^{(m)})$ could be a non-linear function, we can use the SVM (which assumes a linear classifier) to construct classifiers that are nonlinear in the examples of $x_n$. It gives us an advantage to simplify inner product of $\phi(x^{(n)})^T \phi(x^{(m)})$ in the dual SVM. Thus, instead of defining a non-linear feature map and computing the inner product between examples of $\phi(x^{(n)})^T$ and $\phi(x^{(m)})$, we define a function $k(x^{(n)}, x^{(m)})$ between $x^{(n)}$ and $x^{(m)}$. The function $k$ is named kernel function**.**

As a result, in this step (having an inner products), kernel of SVM can help to decrease the processing of inner products (deceasing the practical implementation cost) of $\phi(x^{(n)})^T \phi(x^{(m)})$. We use Gaussian or Radial Basis Function (RBF) to computing of inner products of $\phi(x^{(n)})^T \phi(x^{(m)})$ without any working on mapped data of $\phi(x)$. So, we can show the kernel function as

$$\phi(x^{(n)})^T \phi(x^{(m)}) = k(x^{(n)}, x^{(m)}) \quad (9)$$

where $k$ is RBF kernel is calculated by (10).

$$k(x^{(n)}, x^{(m)}) = exp(-\frac{\| x^{(n)} - x^{(m)} \|^2}{\gamma}) \quad (10)$$

Substituting (10) in (8) yields a new optimization problem that it can be efficiently computed with a cost proportional to the dimension of the input instead of dimension of the features.

$$Max_\alpha \left\{ \sum_{n=1}^N \alpha_n - \frac{1}{2} \sum_{n=1}^N \sum_{m=1}^M \alpha_n \alpha_m y^{(n)} y^{(m)} k(x^{(n)}, x^{(m)}) \right\} \quad (11)$$

Subject to $\quad \sum_{n=1}^N \alpha_n y^{(n)} = 0$

Now, optimization of (11) can be solved to obtain classifying data.

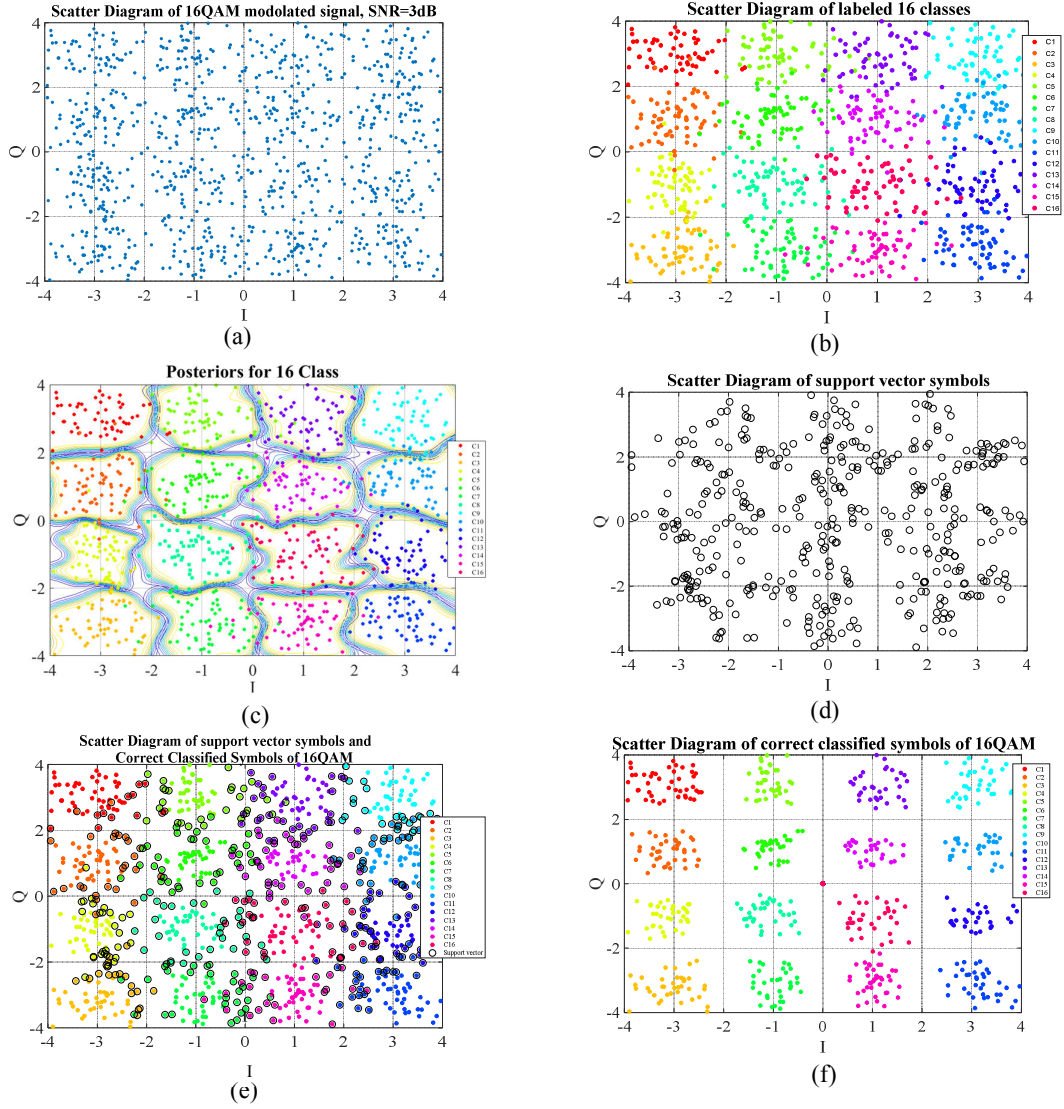

Fig. 2: An example of suggested classifier to 16QAM recognition

$$\hat{y} = sign(w_0 + w^T \phi(x))$$

Where $w = \sum_{\alpha_n > 0} \alpha_n y^{(n)} \phi(x)$

and $w_0 = y^{(s)} - w^T \phi(x^{(s)})$

$$\hat{y} = sign(w_0 + \sum_{\alpha_n > 0} \alpha_n y^{(n)} k(x^{(n)}, x^{(m)}))$$ (12)

$$w_0 = y^{(s)} - w^T \sum_{\alpha_n > 0} \alpha_n y^{(n)} k(x^{(n)}, x^{(m)})$$

Subsequently, Fig. 2 shows how a modulation is recognized by the applied nonlinear soft margin SVM classifier to modulation recognition in the suggested algorithm.

### C. SINR estimation

The SINR measure is computed as a function of the detected support vector symbols in the optimized margin and the received symbols of each signal from SIC. The computed SINR values are mapped to a spectral efficiency measure defined as the product of the number of modulated bits per symbol and the coding rate. For each SINR measure, distinct modulation schemes and coding rates are found through a lookup table, TABLE I.

Let us define the optimum received signal $s_i$ is $i^{th}$ NOMA user from $f^{th}$ subchannel of OFDM. According to Fig. 1 and equation (1), $s_i$ is entered to Machine Learning Unit (SVM) to estimate support vectors in optimum margin. The error signal, including all of support vectors in the margin, can be formulated as $e_i = [e_i^{(v)}]_{1 \times V}$ for $v = 1, 2, \ldots, V$ where $V$ is number of estimated support vectors in the optimized margin. Additionally, the $e_i^{(v)}$ is obtained as follow.

$$e_i^{(v)} = \begin{cases} s_i^{(n)} & if \; y^{(n)}(w^T \phi(x^{(m)}) + w_0) \leq 1 - \xi_n \\ 0 & if \; y^{(n)}(w^T \phi(x^{(m)}) + w_0) > 1 - \xi_n \end{cases} \quad (13)$$

we compute an estimation of the SINR named $Sr$, defined as the ratio of the received signal power $\sigma_{s_i}^2$ to the error signal power $\sigma_{e_i}^2$.

$$Sr = 10 \log_{10}\left(\frac{\sigma_{s_i}^2}{\sigma_{e_i}^2}\right) \quad (14)$$

The support vectors used to SINR estimation can be proportioned or equalized to the method of SINR estimation in the LTE and LTE-Advanced mobile communications standards [70]- [71]. In LTE and LTE-Advanced mobile communications standard, the SINR is defined as the ratio of received signal power to error signal power that received signal should be estimated, and the error signal is taken from decoded bits at the receiver. SINR estimation in LTE standard needs to best estimation of received signal and decoding while suggested method can be well computed by achieved support vectors without any demodulation and decoding. The SINR is calculated frame by frame through the signal transforming. Fig.3 indicates an example of SINR estimation in each frame through connection.

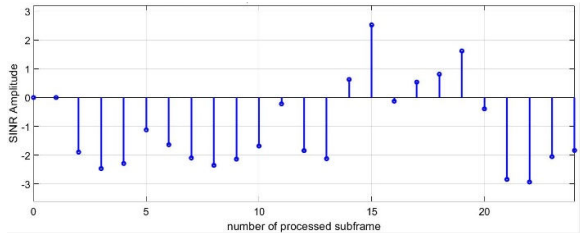

Fig. 3: Frame by frame SINR estimation in the suggested algorithm.

### D. Codding rate estimation

The computed SINR values are mapped to spectral efficiencies indicated in the TABLE I. We also recognized the modulation type of each sub-domain TU-NOMA signals by the aid of SVM classifier. Therefore, according to TABLE I, we be able to estimate codding rate of modulated signals as follow. If $\vartheta_{i_s}$ is the codding rate in the level index $i_s$,

$$\vartheta_{i_s} = \frac{\varpi_{i_s}}{\zeta_{i_m}} \quad (15)$$

where $\varpi_{i_s}$ is the spectral efficiency looked up in TABLE I, and $\zeta_{i_m}$ denotes the number of modulated bits per symbol.

TABLE I : Lookup table for mapping recognized modulation scheme and range of estimated SINR to spectral efficiency

| Level Index ($i_m$ and $i_s$) | Estimated Modulation ($M_{i_m}$) | Bits per symbol ($\zeta_{i_m}$) | SINR (dB) $P_{i_s}^{min} \leq SINR_{i_s} < P_{i_s}^{max}$ | Spectral Efficiency [30] ($\varpi_{i_s}$) |
|---|---|---|---|---|
| 1 | QPSK | 2 | $-6.7 \leq SINR_1 < -4.7$ | 0.1523 |
| 2 | QPSK | 2 | $-4.7 \leq SINR_2 < -2.3$ | 0.2344 |
| 3 | QPSK | 2 | $-2.3 \leq SINR_3 < 0.2$ | 0.3770 |
| 4 | QPSK | 2 | $0.2 \leq SINR_4 < 2.4$ | 0.6016 |
| 5 | QPSK | 2 | $2.4 \leq SINR_5 < 4.3$ | 0.8770 |
| 6 | QPSK | 2 | $4.3 \leq SINR_6 < 5.9$ | 1.1758 |
| 7 | 16QAM | 4 | $5.9 \leq SINR_7 < 8.1$ | 1.4766 |
| 8 | 16QAM | 4 | $8.1 \leq SINR_8 < 10.3$ | 1.9141 |
| 9 | 16QAM | 4 | $10.3 \leq SINR_9 < 11.7$ | 2.4063 |
| 10 | 64QAM | 6 | $11.7 \leq SINR_{10} < 14.1$ | 2.7305 |
| 11 | 64QAM | 6 | $14.1 \leq SINR_{11} < 16.3$ | 3.3223 |
| 12 | 64QAM | 6 | $16.3 \leq SINR_{12} < 18.7$ | 3.9023 |
| 13 | 64QAM | 6 | $18.7 \leq SINR_{13} < 21.0$ | 4.5234 |
| 14 | 64QAM | 6 | $21.0 \leq SINR_{14} < 22.7$ | 4.1152 |
| 15 | 64QAM | 6 | $SINR_{15} \geq 22.7$ | 5.5547 |

### IV. PROPOSED ALGORITHM

In Algorithm. 1, we proposed a ML based algorithm for joint modulation classification and SINR estimation to efficient power control achievement, link adaptation, and capacity enhancement in B5G communication networks. Results of joint estimation of modulation and SINR described previously are used to link adaptation and power control (power reallocation) through the eNode B connections. In other words, link control unit of Fig. 1, in suggested algorithm reassigns types of modulation and codding rate to link adapting, and readjusts power transmit in next frame connection of each signal as power control. In the case of TU-NOMA, after modulation recognition and SINR estimation for both of users, the algorithm compares the results to the looked-up information in TABLE I. In Algorithm. 1, the link adaptation and power control processes are explained. In the beginning of the algorithm, initial data, including example data and their corresponding labels, $X_n$, $Y_n$ respectively, are given to machine learning approach (SVM). In line 3 and 4, the SVM is implemented to joint estimation of modulation type $\widehat{\mathcal{M}}$ and SINR $Sr_{i_s} = \widehat{S}r$. Indexes $i_m$ and $i_s$ indicate the categories of $\widehat{\mathcal{M}}$ and $\widehat{S}r$ in TABLE I. It means $\mathcal{M}_{i_m} = \widehat{\mathcal{M}}$ and $Sr_{i_s} = \widehat{S}r$. The indexes are categorized to three categories according to three desired modulation types used in 5G (accorded to advanced LTE), 1- 6 for QPSK, 7-9 for 16QAM, and 10-15 for 64QAM. In next line, the number of bits per symbol, $\zeta_{i_m}$ belonged to $\widehat{\mathcal{M}}$ is computed by the aid of TABLE I. Afterward, in line 6 to 34, the link evaluation is performed to adaptive modulation, codding, and modifying power in next frame of connection. In this step, depend on $i_m$ and $i_s$ coordination, we totally have three situations as follow.

*1) Coordinated relation between $i_m$ and $i_s$ ($\widehat{\mathcal{M}}$ and $SI\widehat{N}R$).*

This situation happens when $i_m$ and $i_s$ are in the similar category. Additionally, coordinating $i_m$ and $i_s$ means that the physical channel is appropriate to continue with current modulation, codding rate, and allocated power for next frame connection. So, the modulation type is not changed and is adjusted to $\mathcal{M} = \widehat{\mathcal{M}}$. Codding rate is not also changed and is computed from (15). Consequently, power allocation can be modified to minimum power of $i_m$ level as follow.

$$p = p_{i_s}^{min} + \sigma_{e_i}^2 \qquad (16)$$

| | Algorithm. 1: Link adaptation and Power control |
|---|---|
| | Input: 1st user & 2nd user data |
| | Output: Modulation type, SINR, codding rate, transmit power |
| 1 | Initialize $X_n$, $Y_n$ |
| 2 | Repeat |
| 3 | **SVM**: Compute $y, w$, and $w_0$ from equation (12), and Recognize $\widehat{\mathcal{M}}$ (Modulation type). |
| 4 | **SVM**: Compute $SI\widehat{N}R$ from equation (14) and $i_s$ from TABLE I and $\hat{S}r_{i_s} = SI\widehat{N}R$ |
| 5 | Compute $\zeta_{i_s}$ from TABLE I. |
| 6 | **if** $\widehat{\mathcal{M}} = QPSK$ |
| 7 | **if** $1 \leq i_s \leq 6$ |
| 8 | $\mathcal{M} = QPSK$ |
| 9 | **if else** $7 \leq i_s \leq 9$ |
| 10 | $\mathcal{M} = 16QAM$ |
| 11 | **else** $10 \leq i_s \leq 15$ |
| 12 | $\mathcal{M} = 64QAM$ |
| 13 | **end** |
| 14 | $\vartheta = \varpi_{i_s}/\zeta_{i_s}$ from equation (15) |
| 15 | $p = p_{i_s}^{min} + \sigma_{e_i}^2$ from equation (16) |
| 16 | **if else** $\widehat{\mathcal{M}} = 16QAM$ |
| 17 | **if** $1 \leq i_s \leq 6$ |
| 18 | $i_m = 7$ |
| 19 | **if else** $7 \leq i_s \leq 9$ |
| 20 | $\mathcal{M} = 16QAM$ |
| 21 | **else** $10 \leq i_s \leq 15$ |
| 22 | $\mathcal{M} = 64QAM$ |
| 23 | **end** |
| 24 | $\vartheta = \varpi_{i_s}/\zeta_{i_s}$ from equation (15) |
| 25 | $p = p_{i_s}^{min} + \sigma_{e_i}^2$ from equation (16) |
| 26 | **else** $\widehat{\mathcal{M}} = 64QAM$ |
| 27 | **if** $1 \leq i_s \leq 9$ |
| 28 | $i_m = 10$ |
| 29 | **else** $10 \leq i_s \leq 15$ |
| 30 | $\mathcal{M} = 64QAM$ |
| 31 | **end** |
| 32 | $\vartheta = \varpi_{i_s}/\zeta_{i_s}$ from equation (15) |
| 33 | $p = p_{i_s}^{min} + \sigma_{e_i}^2$ from equation (16) |
| 34 | **end** |
| 35 | insert $\Xi = [\mathcal{M}_1, \mathcal{M}_2]$, $\vartheta = [\vartheta_1, \vartheta_2]$, **SINR** $= [\mathcal{S}r_1, \mathcal{S}r_2]$, **P** $= [p_1, p_2]$ to the TU-NOMA. |
| 36 | Until NOMA users |

*2) uncoordinated relation between $i_m$ and $i_s$ with less SINR.*

This situation happens when $i_m > i_s$. For example, $\widehat{\mathcal{M}} = 16QAM$ and $7 \leq i_m \leq 9$ while $\hat{S}r = 2dB$ nad $i_s = 4$. Additionally, uncoordinated relation between $i_m$ and $i_s$ means that the physical channel is not appropriate because the signal have been attenuated through the physical channel. So, the attenuation should be compensated. As a result, the modulation can be unchanged. On the other hand, we can improve the allocated power to compensation of low SINR by means equation (15). If we adjust the allocated power to minimum power that allocated to 16QAM, $i_s=7$, that reserved in TABLE I, we select an efficient power without modulation order decreasing. It means that we establish data rate of 16QAM by power adjusting. After that the efficient power is allocated, codding rate can be correspondingly computed with accorded spectral efficiency in TABLE I and equation (15).

*3) uncoordinated relation between $i_m$ and $i_s$ with more SINR.*

This situation happens when $i_m < i_s$. For example, $\widehat{\mathcal{M}} = 16QAM$, and $7 \leq i_m \leq 9$ while $\hat{S}r = 15dB$, and $i_s = 11$. In simple words, the $\hat{S}r$ is greater than necessary SINR to transmitting a signal that is modulated by $\widehat{\mathcal{M}}$. So, in this case we do not need to any increase in transmit power for the next frame. However, modulation order can be increased to the $\hat{S}r$ level in TABLE I. Moreover, we can decrease the codding rate order because of appropriate SINR level. So, in this case, we be able to increase the network capacity by increase the modulation order and decrease the codding rate. For more demonstration, TABLE II explains the details of motioned three situations in the Algorithm 1.

TABLE II : Link adaptation and power control in three situations as suggested in Algorithm 1.

| Estimated modulation | recognized Modulation Index($i_m$) | Index Estimated SINR($i_s$) | Modulation decision | Codding rate | Power control |
|---|---|---|---|---|---|
| QPSK | $1 \leq i_m \leq 6$ | $1 \leq is \leq 6$ | No change $\mathcal{M} = \widehat{\mathcal{M}}$ | No change from (15) | No change from (16) |
| | | $7 \leq is \leq 9$ | Increase order $\mathcal{M} = 16QAM$ | Increase from (15) | No change from (16) |
| | | $10 \leq is \leq 15$ | Increase order $\mathcal{M} = 64QAM$ | Increase from (15) | No change from (16) |
| 16QAM | $7 \leq i_m \leq 9$ | $1 \leq is \leq 6$ | No change $\mathcal{M} = 16QAM$ | Decrease to $i_s=7$ level from (15) | Increase to $i_s=7$ level from (16) |
| | | $7 \leq is \leq 9$ | No change $\mathcal{M} = 16QAM$ | No change from (15) | No change from (16) |
| | | $10 \leq i_s \leq 15$ | Increase order $\mathcal{M} = 64QAM$ | Increase to $i_s$ level from (15) | No change from (16) |
| 64QAM | $10 \leq i_m \leq 15$ | $1 \leq is \leq 6$ | No change $\mathcal{M} = 64QAM$ | min at $i_s=10$ level from (15) | Increase to $i_s=10$ level from (16) |
| | | $7 \leq is \leq 9$ | No change $\mathcal{M} = 64QAM$ | min at $i_s=10$ level from (15) | Increase to $i_s=10$ level from (16) |
| | | $10 \leq i_s \leq 15$ | No change $\mathcal{M} = 64QAM$ | No change from (15) | Min bound of $i_s$ from (16) |

## V. SIMULATION RESULTS AND ANALYSIS

In this section, simulation results are presented. The results involve Modulation classification success rate, power consumption, capacity, and performance of the 5G communication network that is assisted by SVM classifier.

## A. Modulation classification success rate

One of important criteria to evaluate a modulation classifier is success rate assessment. Success rate presents the reliability of the classification. Machine learning base method to modulation recognition is belonged to FB method. In compare to LB method, the FB classifying approaches, being less complicated and more accurate, are interested in dense user networks. In suggested algorithm, the success rate of modulation classifier has direct effect on the network performance. Moreover, the classifiers depended on SINR value have different rates. They also have different success rates to different modulations and their orders. In this section the success rates of suggested method for the desired modulation QPSK, 16QAM, and 64QAM are reported in Fig. 4.

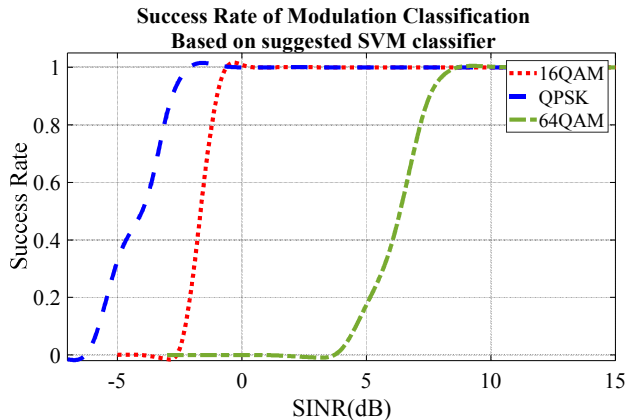

Fig. 4 : Success rate of modulation classification based on SVM in terms of SNR.

Fig. 4 is analyzed in TABLE III. This analysis includes SNR ranges for 99% and 80% success rates. In addition, number of symbols and number of iterations are set $10^3$ and $10^6$, respectively. Where number of symbols indicates the number of symbols that we received from the modulated signal, and iterations are number of the proposed algorithm repetition to classify modulated signal. Additionally, equivalent symbol number provides similar conditions in the modulated signals sampling. In the other hand, the large number of repetition (iteration), $10^6$ for any classification, can improve the reliability of classifier results. Iterations do not have any influences on the success rate, but it can improve the reliability and precise of achieved success rate.

As TABLE III shows, the proposed method achieves 99% success rate at SNR as low as -2.2 dB for QPSK modulations. It also achieves 99% success rate at SNR -1 dB and 8.2 dB to 16QAM and 64QAM, respectively. Moreover, if we desire to 80% success in classifying, the algorithm can recognize QPSK, 16QAM, and 64QAM at -3.2 dB, -1.3dB, and 7dB, respectively.

As a result, in this section we fixed two success rate levels in order to evaluation of the proposed algorithm. We also reported the required SNRs to achieve these success rates for modulation of QPSK, 16QAM, and 64QAM. It is shown that required SNRs are different for the sake of different difficulties in features extraction.

TABLE III: Success rates of the proposed algorithm at different SNRs

| Modulation type | 99% success rate | 80 % success rate | symbols | iteration |
|---|---|---|---|---|
| QPSK | -2.2 dB | -3.2 dB | $10^3$ | $10^6$ |
| 16QAM | -1dB | -1.3 dB | $10^3$ | $10^6$ |
| 64QAM | 8.2 dB | 7 dB | $10^3$ | $10^6$ |

In TABLE IV, we compare our classifier to other classifiers in different SNR, success rates and complexity. For example, success rate of the proposed algorithm for 16-QAM is %99 at SNR=-1 dB, while the best results that have been reported is about SNR=7 dB for ALRT method presented in [72] and for AMC with ELM method presented in [73]. Also, AMC with ELM method uses an extreme learning machine as a classifier, which has a faster learning process and better performance than conventional machine learning method. In addition, in [75] a method based on genetic algorithm has been proposed that needs to SNR=15dB for success rate of 99%. In [76] a feature-based classifier has been suggested that has success rate about 99% with SNR=11dB. Additionally, a method for classifying the electromagnetic signals of a radar or communication system according to their modulation characteristics has been presented in [12]. It identifies 16QAM with success rate of 99.26% and SNR=30 dB.

TABLE IV: Performance comparison with other automatic classifiers.

| Classification algorithm | SNR (dB) | Success rate | Modulation type |
|---|---|---|---|
| Proposed algorithm | -1 | 99% | 16-QAM |
| ALRT[72], L=1,ηA=1 | 7 | 99% | 16-QAM |
| Quasi-ALRT[72] | 30 | 88% | 16-QAM |
| HLRT[72], μ H not specified | 9 | 99% | 16-QAM |
| Cumulant-based[72],Nm=2, μH=−0.68 | 9 | 99% | 16-QAM |
| Quasi-HLRT[72], threshold = 1 | 19 | 99% | 16-QAM |
| AMC with ELM [73] | 7 | 99 % | 16-QAM |
| GPOS [74], symbol length 4096 | 15 | 99% | 16-QAM |
| FB [75] | 11 | 99% | 16-QAM |
| AMC with HMLN [76] | 30 | 99.26% | 16-QAM |

## B. Power consumption

Research on efficient power consumption is a vital issue in next wireless communication network. The algorithms who optimize the consumed power in network are attended to communication system designers. Power control optimization is significantly notable because of at least three reasons. First, less power utilization in network leads to increase user's probability of accessibility to eNode B. Alternately, the suggested method tries to decrease transmitting the unnecessary extra power by the users. In this case, in NOMA system, the extra power can be optimally allocated to more users in each sub-channel of OFDM. It leads to network capacity enhancement. Second, power consuming optimization can also decrease the inter-user interferences in eNode B because it can assist the inter-user power management to avoid transmitting unnecessary extra power by the users through their connections.

Third, the efficient power control can avoid saturation in the receiver of eNode B. For more clearance, Fig. 5 shows the power consumption in the case of suggested algorithm utility in compare of the algorithm that used in advanced LTE [70]- [71]. The result is also compared to an algorithm which selects randomly modulation and codding rate in any frame. All of three algorithms in Fig. 5 implemented to an OFDM-NOMA assisted 5G network.

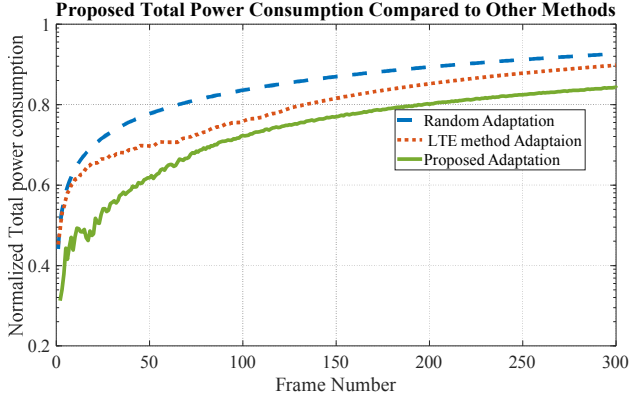

Fig. 5 : Total power consumption in the suggested algorithm compared to other algorithms.

According to whatever Fig. 5 shows, the suggested algorithm can reduce total power consuming through the users' connection.

*C. Capacity improvement*

Assume a TU-NOMA in equation (1), $f^{th} = f$ a fixed frequency for a Sub Channel of OFDM, $h_{f,1} = h_1$, $h_{f,2} = h_2$ the estimated channels, and $s_{f,1} = s_1$, and $s_{f,2} = s_2$ the estimated first and second signal by SIC in Fig. 1 can be written as

$$S = h_1 p_1 s_1 + h_2 p_2 s_2 + \mathcal{N} \quad (17)$$

where $p_1$ and $p_2$ are estimated by equation (16) in Algorithm. 1 and are saved to sequence $\boldsymbol{P} = [p_1, p_2]$. After computing the power and SINR of users, according to [70], achievable data rate of each user is bounded by

$$\mathcal{R}_1 \leq \log_2(1 + \mathcal{S}r_1|h_1 p_1 s_1|^2) \quad (18)$$

$$\mathcal{R}_2 \leq \log_2\left(1 + \frac{\mathcal{S}r_2|h_2 p_2 s_2|^2}{\mathcal{S}r_{12}|h_1 p_1 s_1|^2 + 1}\right) \quad (19)$$

Regarding to inequalities of (18) and (19), the sum channel capacity achieved by TU-NOMA is given by

$$\mathcal{C}_{sum} = c_1 + c_2 = \log_2(1 + \mathcal{S}r_1|h_1 p_1 s_1|^2)$$
$$+ \log_2\left(1 + \frac{\mathcal{S}r_2|h_2 p_2 s_2|^2}{\mathcal{S}r_1 p_1|h_2 p_2 s_2|^2 + 1}\right) \quad (20)$$

In Fig. 6, the achieved sum capacity by means suggested algorithm in compare of mentioned two algorithms, the advanced LTE and random selective scheme, are stated. Results show that the proposed algorithm can achieve larger sum capacity than other methods.

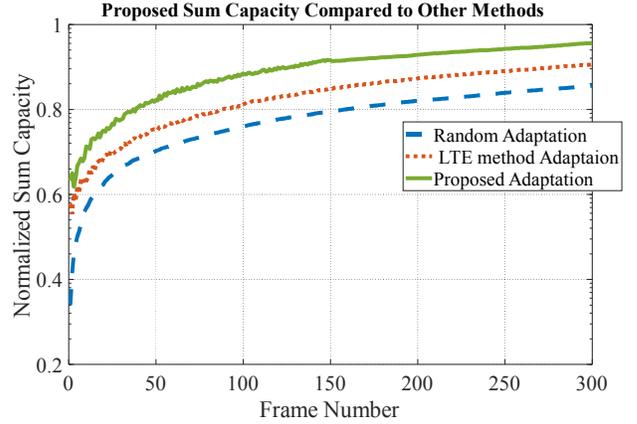

Fig. 6 : Sum capacity in the suggested algorithm compared to other algorithms.

VI. CONCLUSION

In this study, we propose a novel machine learning base algorithm to improve the performance of beyond 5G wireless communication system that is assist by OFDM-NOMA. The suggested method includes an automatic modulation classifier and a SINR estimator that is benefited by non-linear soft margin SVM problem. Obtained support vectors in SVM classifier give us some advantages such as the more accurate modulation recognition and special SINR estimation scheme. In addition, mathematic modelling of AMC and SINR confirm that due to kernel function utility in support vectors computing, the practical implantation is easier than other schemes.

In Fig. 2, SVM problem implemented to modulation recognition (for instance 16QAM) is indicated step by step. The support vectors are computed frame by frame through the enode B connections for each user separately that it excellently can helpful to channel sensing and link adaptation. Link adaptation and power control are performed based on Agoritm.1 and looked up information in TABLE I. Besides, TABLE II briefly explains the Algorithm. 1 construction. For more evidence about proposed algorithm, we evaluated AMC success rate versus SINR, total power consuming, and sum capacity for OFDM-TUNOMA assisted 5G system. Simulation results show an improvement success rate in compare of some published methods. Furthermore, in contrast to other using SINR estimation methods (for example LTE), the submitted SINR estimation do not need to signal decoding. It directly computes SINR after SIC signal detection and before any signal decoding. So, advised SINR estimation is cost-effective method.

Moreover, results prove that the proposed algorithm leads to reducing of total power consumption and increasing of sum capacity through the eNode B connections.

Thus, the presented algorithm can discount occupied symbols to channel quality information in communication signaling because the channel sensing can be done by the algorithm properly.

Finally, a frame by frame power managing scheme and link adapting method are implemented by the aid of a machine learning based modulation classifier and SINR estimator in OFDM-NOMA assisted beyond 5G system. It leads to more success full AMC, efficient SINR estimator, easier practical implantation, overhead signaling discounting, less power consumption, and more capacity achievement.